\documentclass[9pt,technote]{IEEEtran}
\usepackage{CJK}
\usepackage{color,latexsym,amsfonts,amssymb}
\usepackage{amsmath,amsthm}
\usepackage{colordvi,amsfonts,amsmath,epsfig,subfigure}
\usepackage{graphicx}
\input epsf

\def\ba{\begin{array}}
\def\ea{\end{array}}
\def\ban{\begin{eqnarray*}}
\def\ean{\end{eqnarray*}}
\def\bd{\begin{description}}
\def\ed{\end{description}}
\def\be{\begin{equation}}
\def\ee{\end{equation}}
\def\bna{\begin{eqnarray}}
\def\ena{\end{eqnarray}}

\newtheorem{theorem}{\bf Theorem}[section]
\newtheorem{assumption}{\bf Assumption}[section]
\newtheorem{lemma}{\bf Lemma}[section]
\newtheorem{corollary}{\bf Corollary}[section]
\newtheorem{remark}{\bf Remark}

\renewcommand{\thelemma}{\arabic{section}.\arabic{lemma}}

\renewcommand{\thedefinition}{\arabic{section}.\arabic{definition}}

\begin{document}
\title{Multi-Agent Consensus With Relative-State-Dependent Measurement Noises}
\author{Tao Li\thanks{T. Li is with the Key Laboratory of Systems and Control, Academy of
Mathematics and Systems Science, Chinese Academy of Sciences, Beijing 100190, China. Tel.: +86-10-62651440; Fax: +86-10-62587343; Email: {\tt litao@amss.ac.cn}. Tao Li' work is supported by the National Natural Science Foundation of China under grant 61370030.}, Fuke Wu\thanks{Fuke Wu is with School of Mathematics and Statistics, Huazhong
University of Science and Technology, Wuhan 430074, China. Email: {\tt wufuke@mail.hust.edu.cn.
}Fuke Wu's work was supported by the Program for New Century Excellent Talents in University.} and Ji-Feng Zhang\thanks{J. F. Zhang is
with the Key Laboratory of Systems and Control, Academy of Mathematics and Systems Science, Chinese Academy
of Sciences, Beijing 100190, China. Email: {\tt jif@iss.ac.cn}. Ji-Feng Zhang's work is supported by the National Natural Science Foundation of China under grant 61120106011.}
}

\maketitle

\begin{abstract}
In this note, the distributed consensus corrupted by
relative-state-dependent measurement noises is considered. Each agent can measure or receive its neighbors' state information with random noises,
whose intensity is a vector function of agents' relative states.
By investigating the structure of this interaction and the tools of stochastic differential equations,
we develop several small consensus gain theorems to give sufficient conditions in terms of the control gain, the number of agents and the noise intensity function to ensure mean square (m. s.) and almost sure (a. s.) consensus and quantify the convergence rate and the steady-state error. Especially, for the case with homogeneous communication and control channels, a necessary and sufficient condition to ensure m. s. consensus on the control gain is given and it is shown that the control gain is independent of the specific network topology, but only depends on the number of nodes and the noise coefficient constant. For symmetric measurement models, the almost sure convergence rate is estimated by the Iterated Logarithm Law of Brownian motions.
\end{abstract}

\begin{IEEEkeywords}
Multi-Agent system; Distributed coordination; Distributed consensus; Measurement noises; Fading channel.
\end{IEEEkeywords}

\section{INTRODUCTION}

In recent years, the distributed coordination of multi-agent systems with environmental uncertainties has been paid much attention to by the systems and control community. There are various kinds of uncertainties in multi-agent networks,
which have significant influence
on the success of coordination algorithms and performances of the whole network.
For distributed networks, the uncertainties of a single node and link may propagate over the whole network
along with the information exchange among agents. Compared with single-agent systems, the effect of uncertainties of multi-agent
systems on the overall performances is closely related to the pattern of information interaction.
Fruitful results have been achieved for distributed consensus with stochastic disturbances. For discrete-time models, the distributed stochastic approximation method is introduced in \cite{MinyiHuang2006}-\cite{LZ2010TAC} to attenuate the impact of communication/measurement noises
and conditions are given to ensure m. s. and a. s. consensus.
For continuous-time models, Li and Zhang \cite{LZAutomatica2007} gave a necessary and sufficient condition
on the control gain to ensure m. s. consensus.
Wang and Elia \cite{WE2012TAC} made a systematic study of unstable network dynamic behaviors with white Gaussian input noises, channel fading and time-delay. Furthermore, computational expressions for checking m. s. stability under circulant graphs are developed in \cite{WE2013Automatica}. Aysal and Barner \cite{AB2010TSP} and Medvedev \cite{Me2012SICON} studied the distributed consensus with additive random noises for discrete-time and continuous-time models, respectively. In a general framework, Aysal and Barner \cite{AB2010TSP} gave a sufficient condition to ensure a. s. consensus, and Medvedev \cite{Me2012SICON} gave a sufficient condition to ensure closed-loop states to be bounded in m. s..

Most of the above literature assume that
the intensity of noises is time-invariant and independent of agents' states. However, this assumption does not always hold for some
important measurement or communication schemes.
For consensus with quantized measurements (\cite{DiJoAutomatica2010}),
if the logarithmic quantizer (\cite{CFSZ2008Automatica}) is used, then the uncertainties introduced by the quantization are modeled by relative-state-dependent white noises in a stochastic framework (\cite{CFSZ2008Automatica}).
If the relative states are measured by analog fading channels,
the uncertainties of the measurement  are also relative-state-dependent noises (\cite{WE2012TAC}-\cite{WE2013Automatica}, \cite{El2005SCL}).
It is a prominent feature of multi-agent networks with relative-state-dependent noises that
the dynamic evolution of uncertainties of the whole network
interacts with the dynamic evolution of the agents' states in a distributed information architecture, which
results in essential difficulties for the control design and closed-loop analysis of this kind of uncertain multi-agent networks.

In this note, we consider the distributed consensus of high-dimensional first-order agents with
relative-state-dependent measurement noises.
The information interaction of agents is described by
an undirected graph. Each agent can measure or receive its neighbors' state information with random noises.
Different from our previous work for the case with white Gaussian measurement noises (\cite{LZAutomatica2007}), here, the noise intensity
is a vector function of the relative states of the agents. So different from most of the existing literature,
the statistical properties of the impact of the noises on the network are time-varying and coupled by the dynamic evolutions of the agents' states. Since the noise intensity depends on relative states, our model can not be covered by the case with time-varying but independent-of-state noise intensity functions considered in \cite{AB2010TSP}-\cite{Me2012SICON}.
Typical examples for our model are the logarithmic quantization model
in the stochastic framework (\cite{CFSZ2008Automatica}) and the distributed averaging system with Gaussian fading communication channels (\cite{WE2013Automatica}).
We show that the closed-loop consensus error equation becomes a stochastic differential equation with multiplicative noises, which
presents us an interesting property that the coupling is quite well-organized between the noise process over the whole network
and the dynamics of the agents. This equation quantitatively shows how the intensity coefficient matrix associated
with the network noises
relates to the network topology.  For the case with independent and homogeneous control channels and linear noise intensity functions, the quadratic sum of coefficient matrices
over all measurement channels
is exactly the diagonal matrix composed of non-zero eigenvalues of the Laplacian matrix $\mathcal{L}$ multiplied by a constant dependent on
the control gain, the noise intensity coefficient of a single link and the number $N$ of network nodes.
We develop several small consensus gain theorems and show that if the noise intensity function linearly grows with rate bounded by $\overline{\sigma}$, then a control gain $k$ which satisfies $0<k<N/[(N-1)\overline{\sigma}^2]$ can ensure asymptotically unbiased m. s. and a. s. average-consensus, and
the m. s. steady-state error and convergence rate can be given in quantitative relation to the control gain, the noise
and network topology parameters.
Especially, for the case with independent and homogeneous channels, if the noise intensity grows with the rate $\sigma$, then $0<k<N/[(N-1)\sigma^2]$ is also
a necessary and sufficient condition to ensure m. s. consensus.
We show that though a small control gain can decrease the mean-square steady-state error for achieving average-consensus, it may slow down the m. s. convergence rate as well.
For optimizing the m. s. convergence rate, the optimal control gain is $\frac{N}{2(N-1)\sigma^2}$ in some sense.
We prove that for multi-agent networks with relative-state-dependent
measurement noises, the condition for a. s. consensus is weaker than that for m. s. consensus. Especially, for
networks with homogeneous linear growth noise intensity functions and control channels, consensus can be achieved with probability one provided that the control gain satisfies $k+\frac{k^2\sigma^2}{2}>0$. This is a prominent difference compared with the case with non-state-dependent measurement noises (\cite{LZAutomatica2007}). For the case with symmetric noise intensity functions, by the Iterated Logarithm Law of Brownian motions,
it is shown that the convergence rate with probability 1 is between $O(\exp\{(-(k+\frac{k^2\sigma^2}{2})\lambda_{2}(\mathcal{L})+\epsilon)t\})$ and $O(\exp\{(-(k+\frac{k^2\sigma^2}{2})\lambda_{N}(\mathcal{L})-\epsilon)t\})$, $\forall\ \epsilon>0$.


The following notations will be used. $\mathbf{1}$ denotes a column vector with all ones.
$\eta_{N,i}$ denotes the $N$-dimensional column vector with the $i$th element being $1$ and others being zero.
$J_{N}$ denotes the matrix $\frac{1}{N}\mathbf{1}\mathbf{1}^{T}$.
$I_N$ denotes the $N$-dimensional
identity matrix. For a given matrix or vector $A$, $A^T$ denotes its transpose,
and $\|A\|$ denotes its 2-norm.
For two matrices $A$ and $B$, $A\otimes B$ denotes their Kronecker product.
For any given real symmetric matrix $L$, its minimum real eigenvalues is denoted by $\lambda_{\min}(L)$ and the maximum by $\lambda_{\max}(L)$.
For any given square matrix $A$, define $\hat{\lambda}_{\min}(A)=\min_{1\leq i\leq N}\{|\lambda_i(A)|\}$,
and $\hat{\lambda}_{\max}(A)=\max\{|\lambda_i(A)|, \ i=1, 2, \cdots, N\}$.
$\mathbb{E}[\cdot]$ denotes the mathematical expectation.

\section{PROBLEM FORMULATION}\label{sec1}


We consider the consensus control for a
network of the agents with the following dynamics
\bna
\label{agentdynamicequation}
\dot{x}_{i}(t)=u_i(t),\ i=1,2,...,N,
\ena
where $x_{i}(t)\in \mathbb{R}^{n}$ and $u_{i}(t)\in \mathbb{R}^{n}$. Here, each agent has $n$ control channels,
and each component of $x_{i}(t)$ is controlled by a control channel.
Denote $x(t)=[x_1^T(t),...,x_{N}^T(t)]^{T}$ and $u(t)=[u_1^T(t),...,u_{N}^T(t)]^{T}$.
The information flow structure among different agents is modeled as an undirected graph
$\mathcal{G}=\{\mathcal{V}, \mathcal{A}\}$,
where $\mathcal{V}=\{1,2,...,N\}$ is the set of nodes with $i$
representing the $i$th agent, and
$\mathcal{A}$$=$$[a_{ij}]$$\in$$\mathbb{R}^{N\times
N}$ is the adjacency matrix of $\mathcal{G}$ with
element $a_{ij}=1$ or $0$ indicating whether or not there is an
information flow from agent $j$ to agent $i$ directly\footnote{Here, for conciseness, we consider undirected graphs
with $0-1$ weights. It is not difficult to extend our results to the case with general digraphs with  nonnegative weights.}.
Also, $\mbox{deg}_{i}=\sum_{j=1}^{N}a_{ij}$ is called the degree of $i$,
The Laplacian matrix of $\mathcal{G}$ is defined as
$\mathcal{L}=\mathcal{D}-\mathcal{A}$,
where $\mathcal{D}=\mbox{diag}(\mbox{deg}_{1},...,\mbox{deg}_{N})$.
The $i$th agent can receive information from its neighbors with random perturbation as the form:
\bna
\label{measurementequation}
y_{ji}(t)=x_{j}(t)+f_{ji}(x_{j}(t)-x_{i}(t))\xi_{ji}(t),\  j\in N_{i},
\ena
where $N_i=\{j\in \mathcal{V}\ |\ a_{ij}=1\}$ denotes the set of neighbors of agent $i$, $y_{ji}(t)$ denotes the measurement of $x_j(t)$ by agent $i$, and $\xi_{ji}(t)\in\mathbb{R}$ denotes the measurement noise.

\begin{assumption}\label{As1}
The noise intensity function $f_{ji}(\cdot)$ is a mapping from $\mathbb{R}^{n}$ to $\mathbb{R}^{n}$. There exists a constant $\bar{\sigma} \geq 0$ such that
$\|f_{ji}(x)\|\leq\bar{\sigma}\|x\|$,\ $i=1,...,N$,\ $j\in N_{i}$,
for any $x\in\mathbb{R}^n$.
\end{assumption}


\begin{assumption}
\label{As2}
The noise processes $\{\xi_{ji}(t), i,j=1,...,N\}$ satisfy
$\int_{0}^{t}\xi_{ji}(s)ds=w_{ji}(t)$,\ $t\geq0$,
where $\{w_{ji}(t), i,j=1,...,N\}$ are independent Brownian motions.
\end{assumption}

%
%
%
%

\begin{remark}{\rm
Consensus problems with quantized measurements of relative states were studied in \cite{DiJoAutomatica2010}. If the logarithmic quantization is used, then
by properties of  logarithmic quantizers, the quantized measurement by agent $i$ of $x_{j}(t)-x_{i}(t)$ is given by
$z_{ji}(t)=x_{j}(t)-x_{i}(t)+(x_{j}(t)-x_{i}(t))\Delta_{ji}(t)$,
which can be viewed as a special case of (\ref{measurementequation}), where the quantization uncertainty $\Delta_{ji}(t)$ is regarded as white noises (\cite{CFSZ2008Automatica}) in the stochastic framework.
}
\end{remark}

\begin{remark}{\rm
Distributed averaging with Gaussian fading channels were studied in \cite{WE2013Automatica},
where the measurement of $x_{j}(t)-x_{i}(t)$ is given by
$z_{ji}(k)=\xi_{ij}(k)(x_{j}(k)-x_{i}(k))$,
where $\{\xi_{ij}(k)\}$ are independent Gaussian noises with mean value $\mu_{ij}$. Following the method in \cite{El2005SCL},
Wang and Elia (\cite{WE2013Automatica}) transformed the above equation into
$z_{ji}(k)=\mu_{ij}(x_{j}(k)-x_{i}(k))+\Delta_{ij}(k)(x_{j}(k)-x_{i}(k))$,
where $\Delta_{ij}(k)=\xi(k)-\mu_{ij}$ are independent zero-mean Gaussian noises.
This can be viewed as a discrete-time version of (\ref{measurementequation}), where $\mu_{ij}$ can be merged into the weight of the
weighted adjacency matrix of the network topology graph.}
\end{remark}


We consider the following distributed protocol
\bna
\label{controlprotocolexplict}
u_{i}(t)=K\sum_{j=1}^{N}a_{ij}(y_{ji}(t)-x_{i}(t)),\ t\geq0,\ i=1,...,N,
\ena
where $K\in\mathbb{R}^{n\times n}$ is the control gain matrix to be designed.
For the dynamic network (\ref{agentdynamicequation}) and (\ref{measurementequation}) and the distributed protocol  (\ref{controlprotocolexplict}),
we should consider the following questions. (i) Under what conditions is the closed-loop system can achieve m. s. or a. s. consensus? (ii) What is the relationship between the closed-loop performances (i.e. the convergence rate, the steady-state error {\it et al.}) and the control gain matrix $K$, the measurement noise intensity function and the parameters of the network topology graph? How to design the control gain matrix to optimize the closed-loop performances?

\section{MEAN SQUARE AND ALMOST SURE CONSENSUS}

Denote $\delta(t)=[(I_N-J_N)\otimes I_{n}]x(t)$.
Denote $\phi=[\phi_2,...,\phi_{N}]$, where $\phi_{i}$ is the unit eigenvector of $\mathcal{L}$
associated with $\lambda_{i}(\mathcal{L})$. Let $\delta(t)=(T_{\mathcal{L}}\otimes I_{n})\widetilde{\delta}(t)$
and $\widetilde{\delta}(t)=[\widetilde{\delta}_1^{T}(t),...,\widetilde{\delta}_N^{T}(t)]^{T}$ with $\widetilde{\delta}_1(t)\equiv0$. Denote $\overline{\delta}(t)=[\widetilde{\delta}_2^{T}(t),...,\widetilde{\delta}_N^{T}(t)]^{T}$.
Denote $\Lambda^0_{\mathcal{L}}=\mbox{diag}(\lambda_2(\mathcal{L}), \cdots, \lambda_N(\mathcal{L}))$ and
$\Psi_{\mathcal{L}}^{f}(K)=\Lambda_{\mathcal{L}}^0\otimes\left(\frac{K+K^{T}}{2}\right)-\frac{N-1}{N}\|K\|^2\bar{\sigma}^2(\Lambda_{\mathcal{L}}^0\otimes I_n)$, which is a symmetric matrix. We have the following theorem.

\begin{theorem}
\label{sufficientconlemma}
Suppose that Assumptions \ref{As1}-\ref{As2} hold. If $\Psi_{\mathcal{L}}^f(K)$
is positive definite, then the distributed protocol (\ref{controlprotocolexplict}) is an
asymptotically unbiased  m. s. and a. s. average-consensus protocol (\cite{LZAutomatica2007}). Precisely, the closed-loop system
of (\ref{agentdynamicequation}) and (\ref{measurementequation}) under (\ref{controlprotocolexplict}) satisfies:
for any given $x(0) \in\mathbb{R}^{Nn}$, there is a
random vector $x^{*}\in\mathbb{R}^{n}$ with $\mathbb{E}(x^{*})=\frac{1}{N}\sum_{j=1}^{N}x_j(0)$,  such that
$\lim_{t\rightarrow\infty}\mathbb{E}[\|x_{i}(t)-x^{*}\|^2]=0$, and
$\lim_{t\rightarrow\infty}x_{i}(t)=x^{*}$,\ a.s. \ $i=1,...,N$,
and the m. s. steady-state error is given by
\bna
\label{steadystateerroradd1}
\mathbb{E}\Big\|x^*-\frac{1}{N}\sum_{j=1}^{N}x_j(0)\Big\|^2\leq \frac{\|K\|^2\bar{\sigma}^2\lambda_N(\mathcal{L})\|\delta(0)\|^2}{2N^2\lambda_{\min}(\Psi_{\mathcal{L}}^{f}(K))}.
\ena
Moreover, the m. s. convergence rates of $\delta(t)$ is given by
\bna
\label{expoconmeansquareadd1}
\mathbb{E}[\|\delta(t)\|^2]\leq\|\delta(0)\|^2\exp\{-2\lambda_{\min}(\Psi_{\mathcal{L}}^f(K))t\},
\ena
and the a. s. convergence rate is given by
\bna
\label{expoconmeansquareadd2}
\overline{\lim}_{t\to\infty}\frac{\log\|\delta(t)\|}{t}\leq -\lambda_{\min}(\Psi_{\mathcal{L}}^f(K)),\ a.s.
\ena
\end{theorem}

\begin{remark}
{\rm
Generally speaking, the moment exponential stability and the a. s. exponential stability do not imply each other. But under the linear growth conditions on the drift and diffusion terms, the moment exponential stability implies the a. s. exponential stability
(\cite{Mao1997}).
In Section IV, for the case with linear noise intensity functions, one may see that a. s. consensus requires weaker condition than m. s. consensus.}
\end{remark}






\begin{theorem}\textbf{Small consensus gain theorem}:
\label{sufficoraadd1}
Suppose that Assumptions \ref{As1}-\ref{As2} hold. Let the control gain matrix $K=k I_{n}$, where $k\in\mathbb{R}$.
Then the distributed protocol (\ref{controlprotocolexplict}) is an
asymptotically unbiased m. s. and a. s. average-consensus protocol if the graph $\mathcal{G}$ is connected and
$0<k<\frac{N}{(N-1)\overline{\sigma}^2}$.
\end{theorem}

\textbf{Proof}:
From the condition of the theorem, we know that $\Psi_{\mathcal{L}}^{f}(K)=(k\Lambda_{\mathcal{L}}^0-\frac{k^2\bar{\sigma}^2(N-1)}{N}\Lambda_{\mathcal{L}}^0)\otimes I_n$, and so $\Psi_{\mathcal{L}}^{f}(K)$ is positive definite if and only if $\Big(k-\frac{k^2\overline{\sigma}^2(N-1)}{N}\Big)\lambda_{i}(\mathcal{L})>0$, $i=2,...,N$.
The above inequalities hold if and only if the graph $\mathcal{G}$ is connected and $0<k<\frac{N}{(N-1)\overline{\sigma}^2}$.
Then, by
Theorem \ref{sufficientconlemma}, we have the conclusion of the theorem. \qed


\begin{remark}
{\rm Theorem \ref{sufficoraadd1} tells us that for the case with mutually independent and homogeneous control channels, if the graph is connected and the product of  control gain $k$
and the square upper bound of the noise intensity $\overline{\sigma}^2$ is less than $\frac{N}{N-1}$, then both m. s. and a. s. consensus can be achieved. It is obvious that $0<k\overline{\sigma}^2<1$ suffices for $0<k<\frac{N}{(N-1)\overline{\sigma}^2}$, so the selection of
control gain can be independent of $N$ and the network topology, and intuitively speaking,
in inverse proportion to the growth rate of the noise intensity function.
}
\end{remark}

%
%
%

\section{LINEAR NOISE INTENSITY FUNCTION}

In this section, we will consider the
case where the noisy intensity $f_{ji}(\cdot)$ is a linear function of the relative state
$x_{j}(t)-x_{i}(t)$.

\begin{theorem}
\label{MSstability}
Suppose that Assumptions \ref{As1}-\ref{As2} hold with $f_{ji}(x)=\Sigma_{ji}x$,\ $i\neq j$,\ $i,j=1,...,N$,
for any $x\in\mathbb{R}^n$, where $\Sigma_{ji}\in\mathbb{R}^{n\times n}$.
Let $B_{ij}=[b_{kl}]_{N\times N}$ be an $N\times N$ matrix with $b_{ii}=-a_{ij}$, $b_{ij}=a_{ij}$ and
all other elements being zero, $i, j=1,2,...,N$.
Let
$\Phi_{K}=\sum_{i, j=1}^N(\phi^TB^T_{ij}\phi\phi^TB_{ij}\phi)\otimes(\Sigma^T_{ji}K^TK\Sigma_{ji})$
and $\Psi_{K}=\Lambda^0_{\mathcal{L}}\otimes (K+K^{T})-\Phi_K$.
Then if the protocol (\ref{controlprotocolexplict}) is applied to
the system (\ref{agentdynamicequation}) and (\ref{measurementequation}), then the closed-loop system satisfies
\bna
\label{meansquarerate}
\mathbb{E}[\|\delta(t)\|^2]\geq\|\delta(0)\|^2e^{-\lambda_{\max}(\Psi_{K})t},\cr
\mathbb{E}[\|\delta(t)\|^2]\leq\|\delta(0)\|^2e^{-\lambda_{\min}(\Psi_{K})t}.
\ena
If the symmetric matrix $\Psi_{K}$ is positive definite, then the protocol (\ref{controlprotocolexplict}) is an
asymptotically unbiased m. s. and a. s. average-consensus protocol. 
And
\bna
\label{meanconver2add2}
\mathbb{E}\left[\left\|x^{*}-\frac{1}{N}\sum_{j=1}^{N}x_{j}(0)\right\|^2\right]\leq\frac{\lambda_{\max}(\Phi_K)}{N(N-1)\lambda_{\min}(\Psi_K)}\|\delta(0)\|^2,
\ena
where $x^*$ is the limit of $x_i(t)$, $i=1,...,N$, both in m. s. and probability 1.
\end{theorem}



\begin{remark}
{\rm For consensus problems with precise communication, it is always assumed that the states and control inputs of agents are scalars. This
assumption will not loose any generality for the case with precise communication and with non-state-dependent measurement noises, since the state components of the agents are decoupled.
However, for the case with relative-state-dependent measurement noises, from model (\ref{measurementequation}), one may see that the noise intensity
of different state components will be generally coupled together. For the case with linear noise intensity functions, the coupling among
communication channels of different state components means that $\Sigma_{ij}$, $i\neq j$, $i,j=1,...,N$, are not diagonal matrices. From Theorem \ref{MSstability}, one may see that the non-diagonal elements of $\Sigma_{ij}$ indeed have impacts on the consensus conditions and performances.
}
\end{remark}
For the case with decoupled communication channels, we have the following results.
\begin{theorem}\label{MSstability2}
Suppose that  Assumptions \ref{As1}-\ref{As2} hold with
$f_{ji}(x)=\sigma_{ji}x$,\ $\sigma_{ji}>0$,\ $i\neq j$,\ $i, j=1,...,N$, for any $x\in\mathbb{R}^{n}$.
Then the protocol (\ref{controlprotocolexplict}) with $K=kI_{n}$, $k\in\mathbb{R}$, is an
asymptotically unbiased m. s. average-consensus protocol
if the network topology graph $\mathcal{G}$ is connected and
$0<k<\frac{N}{\bar{\sigma}^2(N-1)}$,
and only if the network topology graph $\mathcal{G}$ is connected and
$0<k<\frac{N}{\underline{\sigma}^2(N-1)}$,
where $\bar{\sigma}=\max\{\sigma_{ji}, i=1,...,N, j\in N_{i}\}$ and $\underline{\sigma}=\min\{\sigma_{ji}, i=1,...,N, j\in N_{i}\}.$
\end{theorem}



\begin{corollary}\label{cor1}
Suppose that Assumptions \ref{As1}-\ref{As2} hold with
$f_{ji}(x)=\sigma x$\ $i\neq j$,\ $i,j=1,...,N$, for any $x\in\mathbb{R}^n$, where $\sigma>0$.
Then the protocol (\ref{controlprotocolexplict}) with $K=kI_{n}$, $k\in\mathbb{R}$, is an
asymptotically unbiased mean-square average-consensus protocol
if and only if the network topology graph $\mathcal{G}$ is connected and
$0<k<\frac{N}{\sigma^2(N-1)}$.
\end{corollary}




\begin{remark}
{\rm Theorems  \ref{MSstability2} and Corollary \ref{cor1} are concerned with the case where the communication and control channels for
different components of the states of agents are completely decoupled. Especially, in Corollary \ref{cor1}, when the
noise intensity functions are homogeneous for different agents and state components,  we give a necessary and
sufficient condition on the control gain, the noise intensity and network parameters to ensure m. s. consensus. Theorem \ref{sufficoraadd1} showes that if the noise intensity function grows linearly with rate bounded by $\overline{\sigma}$,
then a positive control gain $k<1/\overline{\sigma}^2$ is sufficient for m. s. consensus. For the case of  Corollary  \ref{cor1},
we can see that it is necessary for m. s. consensus
that the upper bound of the control gain is inversely proportional to the square of the growth rate of the noise intensity function.}
\end{remark}


\begin{remark}
\label{remarkcomputeaddadd1}
{\rm From (\ref{meansquarerate}), we  can see that for the case with linear noise intensity functions, the
m. s. convergence rate is controlled by the maximal and minimal eigenvalues of $\Psi(K)$. A question is whether
we can choose $K$ to maximize the m. s. convergence rate. Generally speaking,
a given control gain $K$ that maximize $\lambda_{\min}(\Psi(K))$ may not maximize $\lambda_{\max}(\Psi(K))$ in the meanwhile.
However, Corollary \ref{cor1} tells us that for the case with independent and homogeneous communication and control channels, we can indeed get some optimal solution
of the control gain.
Noting that $\Sigma_{ij}=\sigma I_{n}$, $i, j=1,...,N$, we have
$\Phi_K
=\frac{2(N-1)\sigma^2k^2}{N}\Lambda^0_{\mathcal{L}}\otimes I_n$,
and
$\Psi_K
=\Big(2k-\frac{2(N-1)\sigma^2k^2}{N}\Big)(\Lambda^0_{\mathcal{L}}\otimes I_n)$.
For this case, the eigenvalues of $\Psi_K$ are just the nonzero eigenvalues of the Laplacian matrix multiplied by $2k-\frac{2(N-1)\sigma^2k^2}{N}$.
Let $K^*=\frac{N}{2(N-1)\sigma^2}I_{n}$, then
$\Psi_{K^*}=\max_{K=kI_{n}, 0<k<\frac{N}{\sigma^2(N-1)}}\Psi_{K}$.
This implies that the control gain to optimize
the m. s. convergence rate can be selected as
$k^*=\frac{N}{2(N-1)\sigma^2}$.
}
\end{remark}


\begin{remark}
{\rm From (\ref{meanconver2add2}), we can see that the m. s. steady-state error for average-consensus is bounded by $\frac{\lambda_{\max}(\Phi_K)}{N(N-1)\lambda_{\min}(\Psi_K)}\|\delta(0)\|^2$. The coefficient of the bound
depends on the control gain and the network topology. For the
case of  Corollary  \ref{cor1}, by Remark \ref{remarkcomputeaddadd1}, it can be computed that
$\frac{\lambda_{\max}(\Phi_K)}{N(N-1)\lambda_{\min}(\Psi_K)}=\frac{\sigma^2k\lambda_{N}(\mathcal{L})}{N(N-(N-1)\sigma^2k)\lambda_2(\mathcal{L})}$,
which vanishes as $k\sigma^2\to0$. To reduce the steady-state error  for average-consensus, one way is to decrease the control gain $k$, however, from (\ref{meansquarerate}), we can see that as $k\to0$, the convergence will become very slow; the other way is to
design the network topology to maximize the synchronizability of the network $\lambda_{2}(\mathcal{L})/\lambda_{N}(\mathcal{L})$.
}
\end{remark}

%
\begin{remark}
{\rm 
For the asymptotic analysis, we consider
a sequence $\{\mathcal{G}_{N}, N\geq1\}$ of connected graphs.
Noting that $\lambda_{N}(\mathcal{L})\leq2d(\mathcal{G}_{N})$ and $\lambda_{2}(\mathcal{L})\geq\frac{4}{diam(\mathcal{G}_{N})}$ (\cite{Ab2007LAA}),
we have
$\mathbb{E}\|x^{*}-\frac{1}{N}\sum_{j=1}^{N}x_{j}(0)\|^2$ $\leq$
$\frac{\sigma^2kd(\mathcal{G}_{N})(N-1)}{2N^2(1-\frac{N-1}{N}\sigma^2k)}$,
where $d(\mathcal{G}_{N})$ is the degree of $\mathcal{G}_{N}$ and $diam(\mathcal{G}_{N})$ is the diameter of $\mathcal{G}_{N}$.\footnote{The distance between two vertices in a graph is the length of (i.e., number of edges in) the shortest path between them. The diameter of a graph $\mathcal{G}$  is
maximum distance between any two vertices of $\mathcal{G}$.}
}
\end{remark}

Similar to Theorem \ref{MSstability}, the conditions of Theorem \ref{MSstability2} and Corollary \ref{cor1} suffice for a. s. consensus.  It was shown that for the case with non-state-dependent measurement noises, the conditions for a. s. consensus are the same as those for m. s. consensus (\cite{LZAutomatica2007}, \cite{WZ2009JSSMS}).  From the following theorems, we can see that for the case with relative-state-dependent measurement noises, a. s. consensus requires weaker condition on
the control gain than m. s. consensus.

\begin{theorem}\label{Linalmoststability}
Let
$\mu\doteq\inf_{x\in\mathbb{R}^{n(N-1)}, x\neq 0}\Big\{\frac{1}{\|x\|^2}x^T\Psi_K x+\frac{2}{\|x\|^4}\sum_{i, j=1}^N[x^T((\phi^TB_{ij}\phi)\otimes(K\Sigma_{ji}))x]^2\Big\}$.
Under the assumptions of Theorem \ref{MSstability},  if  the protocol (\ref{controlprotocolexplict}) is applied to
the system (\ref{agentdynamicequation}) and (\ref{measurementequation}), then the closed-loop system satisfies
$\overline{\lim}_{t\to\infty}\frac{\log\|\delta(t)\|}{t}\leq-\frac{\mu}{2}$  a.s.
Particularly, the protocol (\ref{controlprotocolexplict}) is an asymptotically unbiased a. s. average-consensus protocol if $\mu>0$.
\end{theorem}


%
\begin{remark}
{\rm Theorem \ref{MSstability} showes that if $\Psi_{K}$ is positive definite, then the protocol (\ref{controlprotocolexplict})
can drive the dynamic network to consensus both in m. s. and probability $1$. We know that $\mu>0$ is weaker than the positive definiteness of $\Psi_K$ since $2\sum_{i,j=1}^N[x^T((\phi^TB_{ij}\phi)\otimes(K\Sigma_{ji}))x]^2/\|x\|^4>0$ for any $x\neq 0$. This implies that $\mu\geq\lambda_{\min}(\Psi_K)$. Actually, let
$\lambda_K\doteq\lambda_{\min}(\Psi_K)+\frac{1}{2}\sum_{i,j=1}^N\hat{\lambda}^2_{\min}((\phi^TB_{ij}\phi)\otimes(K\Sigma_{ji})+(\phi^TB^T_{ij}\phi)\otimes(\Sigma^T_{ji}K^T))$.
It follows that $\mu\geq\lambda_K$ and $\lambda_K>0$ if $\Psi_K$ is positive definite. So, $\lambda_K>0$ can be used as a sufficient condition, which is easier to be verified than $\mu>0$, to ensure a. s. consensus.
If $\lambda_K>0$, then the closed-loop system satisfies
$\overline{\lim}_{t\to\infty}\frac{\log \|\delta(t)\|}{t}\leq-\frac{\lambda_K}{2}<0$ a.s.
}
\end{remark}

If measurement model is symmetric and $K$ is a symmetric matrix, then more precise estimates of the convergence rate for a. s. consensus can be obtained.

\begin{assumption}
\label{As7}
The noise processes $\{\xi_{ji}(t), i, j=1,...,N\}$ satisfy
$\int_{0}^{t}\xi_{ji}(s)ds=w_{ji}(t)$, $w_{ji}(t)\equiv w_{ij}(t)$, $t\geq0$,
where $\{w_{ji}(t), i=1,...,N, j=1,2...,i\}$ are independent Brownian motions.
\end{assumption}


\begin{theorem}
\label{symmetricconvergencerate}
Suppose that Assumptions \ref{As1} and \ref{As7} hold with $f_{ji}(x)=\sigma_{ji}x$\ $i\neq j$,\ $i,j=1,...,N$,
for any $x\in\mathbb{R}^n$,  where $\sigma_{ji}=\sigma_{ij}>0$.
Apply the protocol (\ref{controlprotocolexplict})
to the system (\ref{agentdynamicequation}) and (\ref{measurementequation}). If  $K$ is symmetric, then the closed-loop system satisfies
\bna
\label{upperboundrate}
\overline{\lim}_{t\to\infty}\frac{\log\|\delta(t)\|+\lambda_{\min}(A_{\mathcal{L}}(K))t}{\sqrt{2t\log\log t}}\leq\hat{\lambda}_{\max}(B_{\mathcal{L},K}),\ a.s.,
\ena
and
\bna
\label{lowerboundrateadd2}
\underline{\lim}_{t\to\infty}\frac{\log\|\delta(t)\|+\lambda_{\max}(A_{\mathcal{L}}(K))t}{\sqrt{2t\log\log t}}\geq-\hat{\lambda}_{\max}(B_{\mathcal{L},K}),\ a.s.,
\ena
where
$A_{\mathcal{L}}(K)=\big[\Lambda_{\mathcal{L}}^0\otimes K+\frac{1}{2}\big(\phi^{T}(\sum_{i,j=1}^NB^2_{ij}\sigma_{ji}^2)\phi\big)\otimes K^2
\big]$, and
$B_{\mathcal{L},K}=\big(\phi^{T}(\sum_{i,j=1}^{N}B_{ij}\sigma_{ji})\phi\big)\otimes K$.
\end{theorem}


\vskip 0.2cm

\begin{corollary}\label{symmetricalmostsurecorollary}
Suppose that the network topology graph $\mathcal{G}$ is connected  and Assumptions \ref{As1} and \ref{As7} hold with $f_{ji}(x)=\sigma x$\ $i\neq j$,\ $i,j=1,...,N$, for any $x\in\mathbb{R}^n$, where $\sigma>0$.
Then the protocol (\ref{controlprotocolexplict}) with $K=kI_{n}$ is an
asymptotically unbiased a. s. average-consensus protocol if $k+\frac{k^2\sigma^2}{2}>0$ and
the convergence rate is given by
\ban
\overline{\lim}_{t\to\infty}\frac{\log\|\delta(t)\|+(k+\frac{k^2\sigma^2}{2})
\lambda_{2}(\mathcal{L})t}{\sqrt{2t\log\log t}}\leq |k|\sigma\lambda_N(\mathcal{L}),\ a.s.,
\ean
and
\ban
\underline{\lim}_{t\to\infty}\frac{\log\|\delta(t)\|+(k+\frac{k^2\sigma^2}{2})
\lambda_{N}(\mathcal{L})t}{\sqrt{2t\log\log t}}\geq -|k|\sigma\lambda_N(\mathcal{L}),\ a.s.
\ean
\end{corollary}

\textbf{Proof}: From $B_{ij}^{2}=-B_{ij}$ and $\sum_{i,j=1}^{N}B_{ij}=-\mathcal{L}$,
 we have $A_{\mathcal{L}}(K)=(k+\frac{k^2\sigma^2}{2})(\Lambda_{\mathcal{L}}^0\otimes I_{n})$ and $B_{\mathcal{L},K}=-k\sigma(\Lambda_{\mathcal{L}}^0\otimes I_{n})$. Then the conclusion follows from Theorem \ref{symmetricconvergencerate}. \qed



\begin{remark} {\rm Corollary \ref{symmetricalmostsurecorollary} tells us that provided that the network is connected, any given positive control gain or negative control gain satisfying $\frac{k\sigma^2}{2}<-1$ can ensure a. s. consensus. Corollary \ref{cor1} tell us that to ensure m. s. consensus, the control gain has to be positive and
small enough such that $\frac{k\sigma^2(N-1)}{N}<1$. This implies that for the case with homogeneous communication and control channels, a. s. consensus require weaker condition than m. s.
consensus, which is consistent with Theorems \ref{MSstability} and \ref{Linalmoststability}.}
\end{remark}

\begin{remark} {\rm For the consensus system with precise communication:
$\dot{x}_{i}(t)=k\sum_{j\in N_{i}}(x_{j}(t)-x_{i}(t))$,
it was shown in \cite{Olfati-Saber2004} that a necessary and sufficient condition on the control gain $k$ for consensus to be achieved is $k>0$.
In \cite{LZAutomatica2007}, for the consensus system with non-state-dependent additive noise:
$\dot{x}_{i}(t)=k\sum_{j\in N_{i}}(x_{j}(t)-x_{i}(t)+\xi_{ji}(t))$,
it was shown that a constant control gain $k$, no matter how small it is, can not ensure the closed-loop stability.
For the consensus system with non-state-dependent measurement noises and the stochastic approximation type control protocol:
$\dot{x}_{i}(t)=k(t)\sum_{j\in N_{i}}(x_{j}(t)-x_{i}(t)+\xi_{ji}(t))$,
it was shown in \cite{LZAutomatica2007} and \cite{WZ2009JSSMS} that the necessary and sufficient condition on the nonnegative control gain $k(t)$ for consensus to be achieved almost surely is $\int_{0}^{\infty}k(t)=\infty$ and $\int_{0}^{\infty}k^2(t)<\infty$.
Corollary \ref{symmetricalmostsurecorollary} tells us that for the consensus system with relative-state-dependent measurement noises:
$\dot{x}_{i}(t)=k\sum_{j\in N_{i}}(x_{j}(t)-x_{i}(t)+(x_{j}(t)-x_{i}(t))\xi_{ji}(t))$,
a sufficient condition on the control gain $k$ for consensus to be achieved almost surely is $k+\frac{k^2\sigma^2}{2}>0$, which means that
even a negative control gain may ensure consensus as well. This tells us that differently from the non-state-dependent measurement noises (\cite{LZAutomatica2007}, \cite{WZ2009JSSMS}), the relative-state-dependent measurement noises will sometimes be helpful for the a. s. consensus of the network. Whether or not network noises need to be attenuated depends on the pattern that noises impact on the network.}
\end{remark}


\begin{remark}
{\rm For the consensus system with non-state-dependent measurement noises and the stochastic approximation type control protocol,
it was shown in \cite{LZAutomatica2007} and \cite{WZ2009JSSMS} that the vanishing control gain $k(t)$  with a proper vanishing speed is necessary and sufficient to ensure the m. s. and a. s. consensus, however, the vanishing control gain may result in a slower convergence of the closed-loop system, which is no longer exponentially fast. From the results of this paper, we can see that for the case with relative-state-dependent
noises, the vanishing of the control gain is not necessary and the convergence speed of the closed-loop system can be exponentially fast.}
\end{remark}


\begin{remark}
{\rm It is well known that multiplicative noises can be used to stabilize an unstable system in the sense of probability 1 (\cite{Wu2009}),
and $p-$moments with $p\in(0, 1)$ (\cite{Zong2013}).
In Corollary \ref{symmetricalmostsurecorollary}, the condition $k+\frac{k^2\sigma^2}{2}>0$ shows that the noises play positive roles for a. s. consensus. However, For the m. s. consensus ($p=2$), the condition $0<\frac{k\sigma^2(N-1)}{N}<1$
in Corollary \ref{cor1} shows that the noises play negative roles, which means that for a given fixed update gain, the noise level $\sigma^2$ could not be larger than the threshold value $\frac{N}{k(N-1)}$. This implies that there exist fundamental differences between the a. s. and the m. s. consensus for the consensus system with relative-state-dependent noises. }
\end{remark}

\begin{remark}
{\rm In \cite{WE2013Automatica}, the discrete-time distributed averaging is considered with fading channels and time-delays. By converting
the fading channel model into a precise measurement model with relative-state-dependent noises and the assumption that the closed-loop system is input-output stable, some necessary and sufficient conditions were given under circulant graphs. It was shown that as the number of agents or time-delay increases to infinity,  small control gains can lead to m. s. stability but may slow down the convergence. Here, the closed-loop system is not assumed to be input-output stable in prior. The network topology is not limited to circulant graphs and the convergence and performance are also considered for a. s. convergence.  Corollary \ref{cor1} shows that to ensure m. s. consensus, the control gain has to be small enough.  This result and those of \cite{WE2013Automatica} both reveal that there is a natural trade-off between the m. s. convergence speed and the robustness to noise for the choice of the control gain. Our method can be further extended to the discrete-time case with the noises modeled by martingale difference sequences, which can cover the case with Bernoulli fading channels (\cite{WE2012TAC}) and the stochastic logarithmic quantization in \cite{CFSZ2008Automatica}.}
\end{remark}

\begin{remark}
{\rm Consider a connected two-agent undirected network. if $f_{12}(x)=\sigma_{12}x,\ f_{21}(x)=\sigma_{21}x$, $\sigma_{12}>0$, $\sigma_{21}>0$, for any $x\in\mathbb{R}^n$, then the protocol (\ref{controlprotocolexplict}) with $K=kI_{n}$, $k\in\mathbb{R}$ is an a. s. average-consensus protocol if and only if $2k+\frac{k^2}{2}(\sigma_{12}^2+\sigma_{21}^2)>0$, and is a m. s.
average-consensus protocol if and only if $4k-k^2(\sigma_{12}^2+\sigma_{21}^2)>0$.
The m. s. steady-state error is given by
$E\|x^*-\frac{x_1(0)+x_2(0)}{2}\|^2=\frac{k(\sigma^2_{12}+\sigma^2_{21})\|x_1(0)-x_2(0)\|^2}{4[4-k(\sigma^2_{12}+\sigma^2_{21})]}$.
So the fact that a. s. requires weaker condition than m. s. consensus can be further verified by the two agent case even
if the channel is not symmetric. It can be verified that
for the two agent case, $\frac{\lambda_{\max}(\Phi_K)}{N(N-1)\lambda_{\min}(\Psi_K)}\|\delta(0)\|^2
=\frac{k(\sigma^2_{12}+\sigma^2_{21})\|x_1(0)-x_2(0)\|^2}{4[4-k(\sigma^2_{12}+\sigma^2_{21})]}$, which implies that the upper bound of the m. s. steady-state error in Theorem \ref{MSstability} is tight for the two agent case.}
\end{remark}


\section{CONCLUDING REMARKS}

In this note, the distributed consensus of high-dimensional first-order agents with
relative-state-dependent measurement noises has been considered. The information exchange among agents is described by
an undirected graph. Each agent can measure or receive its neighbors' state information with random noises,
whose intensity is a vector function of agents' relative states.
By investigating the structure of the interaction between network noises and the agents' states and the tools of stochastic differential equations, we have developed several small consensus gain theorems to give sufficient conditions to ensure m. s. and a. s. consensus and quantify the convergence rate and the steady-state error. Especially, for the case with
linear noise intensity functions and homogeneous communication and control channels, a necessary and sufficient condition to ensure m. s. consensus on the control gain $k$ is $0<k<N/[(N-1)\sigma^2]$, where $\sigma$ is the growth rate of the noise intensity function.
It is shown that for this kind of multi-agent networks, a. s. consensus requires weaker conditions than m. s. consensus.
 Especially, for networks with homogeneous linear noise intensity functions and control channels, consensus can be achieved with probability one provided $k+\frac{k^2\sigma^2}{2}>0$, which means that even a negative control gain can also ensure almost consensus.
For future research on the distributed coordination of multi-agent systems with relative-state-dependent measurement noises, there are many interesting topics, such as the discrete-time case with the noises modeled by martingale difference sequences, the case with random link failures, the time-delay and distributed tracking problems.

\section*{Appendix}
\setcounter{lemma}{0}
\def\thelemma{A.\arabic{lemma}}
\setcounter{definition}{0}
\def\thedefinition{A.\arabic{definition}}
\setcounter{equation}{0}
\def\theequation{A.\arabic{equation}}

\begin{lemma}
\label{philemma1}
The $N\times (N-1)$ dimensional matrix $\phi$ satisfies
$\phi\phi^{T}=I_{N}-J_{N}$, and $\phi^{T}\phi=I_{N-1}$.
\end{lemma}
%
%
%
\begin{lemma}
\label{bijtechlemmaadd1}
Let $B_{ij}=[b_{kl}]_{N\times N}$, $i, j=1,2,...,N$ be matrices defined in Theorem \ref{MSstability}. Then
$B^T_{ij}\mathbf{1}\mathbf{1}^TB_{ij}=\frac{N}{N-1}B^T_{ij}\phi\phi^T B_{ij}$.
\end{lemma}
%

\begin{lemma}
\label{linearclosedloopsystemadd1}
Suppose that the assumptions of Theorem \ref{MSstability} hold. Applying the protocol (\ref{controlprotocolexplict}) to
the system (\ref{agentdynamicequation}) and (\ref{measurementequation}), the closed-loop system satisfies
$d\overline{\delta}(t)
=-(\Lambda^0_{\mathcal{L}}\otimes K)
\overline{\delta}(t)dt+\sum_{i,j=1}^{N}
[(\phi^{T}B_{ij}\phi)\otimes(K\Sigma_{ji})]
\overline{\delta}(t)
dw_{ji}(t)$.
\end{lemma}

\begin{lemma}\label{ORIG}
Suppose that the assumptions of Theorem \ref{MSstability} hold. Apply the protocol (\ref{controlprotocolexplict}) to the system (\ref{agentdynamicequation}) and (\ref{measurementequation}), then for all $\overline{\delta}(0)\neq0$, the closed-loop system satisfies $\mathbb{P}\{\overline{\delta}(t)\neq0\ \mbox{on all }\ t\geq 0\}=1$.
\end{lemma}


\begin{lemma}
\label{consensuerrorexplicitsolutionsymmetric}
Suppose that the assumptions of Theorem \ref{symmetricconvergencerate} hold. Apply the protocol (\ref{controlprotocolexplict}) to the system (\ref{agentdynamicequation}) and (\ref{measurementequation}), then the closed-loop system satisfies
$\overline{\delta}(t)=\exp\left\{-A_{\mathcal{L}}(K)t+M_{\mathcal{L},K}(t)\right\}\overline{\delta}(0)$,
where
$A_{\mathcal{L}}(K)=\left[\Lambda^0_{\mathcal{L}}\otimes K+\frac{1}{2}\sum_{i, j=1}^N[(\phi^{T}B^2_{ij}\phi)\otimes(K\sigma_{ji})^2]
\right]$,
and
$M_{\mathcal{L},K}(t)=\sum_{i,j=1}^{N}[(\phi^{T}B_{ij}\phi)$ $\otimes(K\sigma_{ji})]$ $w_{ji}(t)$.
\end{lemma}


The proofs of Lemmas are omitted here.

\textbf{Proof of Theorem \ref{sufficientconlemma}}:
Substituting the protocol (\ref{controlprotocolexplict}) into the system (\ref{agentdynamicequation}) gives
$\dot{x}_i(t)=K\sum_{j=1}^Na_{ij}(x_j(t)-x_i(t))+K\sum_{j=1}^{N}a_{ij}f_{ji}(\delta_j(t)-\delta_i(t))\xi_{ji}(t)$.
By Assumption \ref{As2}, we have
$dx(t)=-(\mathcal{L}\otimes K)x(t)dt+\sum_{i=1}^{N}\sum_{j=1}^{N}a_{ij}[\eta_{N,i}\otimes(Kf_{ji}(\delta_j(t)-\delta_i(t)))]dw_{ji}(t)$,
which together with the definition of $\delta(t)$ gives
\ban
d\delta(t)&=&-(\mathcal{L}\otimes K)\delta(t)dt\cr
&&+\sum_{i,j=1}^{N}a_{ij}[(I_N-J_N)\eta_{N,i}\otimes(Kf_{ji}(\delta_j(t)\cr
&&-\delta_i(t)))]dw_{ji}(t).
\ean
Then by the definition of $\bar{\delta}(t)$, we have
\ban
d\overline{\delta}(t)&=&-(\Lambda_{\mathcal{L}}^0\otimes K)\overline{\delta}(t)dt\cr
&&+\sum_{i,j=1}^{N}a_{ij}[\phi^{T}(I_N-J_N)\eta_{N,i}\cr
&&\otimes(Kf_{ji}(\delta_j(t)-\delta_i(t)))]dw_{ji}(t).
\ean
By the definitions of $\eta_{N, i}$ and $J_N$, we have
$\eta^T_{N, i}(I_N-J_N)\eta_{N, i}=\frac{N-1}{N}$.
By  Lemma \ref{philemma1}, noting that $(I_N-J_N)^2=I_N-J_N$, applying the It\^{o} formula to $\|\overline{\delta}(t)\|^2$, we get
\ban
d\|\overline{\delta}(t)\|^2&=&-\overline{\delta}^T(t)(\Lambda_{\mathcal{L}}^0\otimes(K+K^{T}))\overline{\delta}(t)dt+dM_1(t)\cr
&&+\frac{N-1}{N}\sum_{i=1}^N\sum_{j=1}^Na^2_{ij}(f^T_{ji}(\delta_j(t)\cr
&&-\delta_i(t))K^TKf_{ji}(\delta_j(t)-\delta_i(t)))dt,
\ean
where $dM_1(t)=2\sum_{i,j=1}^N\overline{\delta}^T(t)a_{ij}[\phi^{T}(I_N-J_N)\eta_{N,i}\otimes(Kf_{ji}(\delta_j(t)-\delta_i(t)))]dw_{ji}(t)$.
By Assumption \ref{As1}, we have
$d\|\overline{\delta}(t)\|^2\leq-2\lambda_{\min}(\Psi^f_{\mathcal{L}}(K))\|\bar{\delta}(t)\|^2$ $+dM_1(t)$,
Then by
the comparison theorem  (\cite{MM1977}), we get (\ref{expoconmeansquareadd1}), which together with the positive definiteness of $\Psi_{\mathcal{L}}^{f}(K)$ leads to
$\lim_{t\to\infty}\mathbb{E}[\|\delta(t)\|^2]=0$.
By the properties of the matrix $\mathcal{L}$, we have
\ban
(\mathbf{1}^{T}\otimes I_n)x(t)=(\mathbf{1}^{T}\otimes I_n)x(0)+\sum_{i,j=1}^{N}a_{ij}M_{ij}(t),
\ean
where $M_{ij}(t)=\int_{0}^{t}[\mathbf{1}^{T}\eta_{N,i}\otimes(Kf_{ji}(\delta_j(s)-\delta_i(s)))]dw_{ji}(s)$.
By Assumption \ref{As1}, noting that $\mathbf{1}^T\eta_{N, i}=1$, it is estimated that
\ban
\mathbb{E}[\int_{0}^{t}\|\mathbf{1}^{T}\eta_{N,i}\otimes(Kf_{ji}(\delta_j(s)-\delta_i(s)))\|^2ds]
\leq\frac{\|K\|^2\bar{\sigma}^2\|\delta(0)\|^2}{\lambda_{\min}(\Psi_{\mathcal{L}}^{f}(K))},
\ean
which implies that $M_{ij}(t)$ is a square-integrable continuous martingale. Then we know that
as $t\to\infty$, $\frac{1}{N}(\mathbf{1}^{T}\otimes I_n)x(t)$ converges to a random variable with finite second-order moment both in mean square and almost surely. Denote the limit random variable by
$x^{*}=\frac{1}{N}(\mathbf{1}^{T}\otimes I_n)x(0)+\frac{1}{N}\sum_{i,j=1}^{N}a_{ij}\int_{0}^{\infty}Kf_{ji}(\delta_j(t)-\delta_i(t))dw_{ji}(t)$
with $E(x^{*})=\frac{1}{N}\sum_{j=1}^{N}x_j(0)$. This together with the convergence of $\mathbb{E}[\|\delta(t)\|^2]$  means
that (\ref{controlprotocolexplict}) is an asymptotically unbiased m. s. average-consensus protocol.
By the definition of $x^*$, we have
\ban
&&\mathbb{E}\Big[\|x^*-\frac{1}{N}\sum_{j=1}^{N}x_j(0)\|^2\Big]\cr
&\leq&\frac{\|K\|^2\bar{\sigma}^2}{N^2}\mathbb{E}\int^{\infty}_0\sum_{i=1}^N\sum_{j=1}^Na_{ij}\|\delta_j(s)-\delta_i(s)\|^2ds\cr
&=&\frac{2\|K\|^2\bar{\sigma}^2}{N^2}\mathbb{E}\int^{\infty}_0\bar{\delta}^T(s)(\Lambda_{\mathcal{L}}^0\otimes I_n)\bar{\delta}(s)ds\cr
&\leq&\frac{\|K\|^2\bar{\sigma}^2\lambda_N(\mathcal{L})\|\delta(0)\|^2}{N^2\lambda_{\min}(\Psi_{\mathcal{L}}^{f}(K))},
\ean
which gives the steady-state error (\ref{steadystateerroradd1}).
It is known that there exists a positive constant $\alpha_1$, such that
$\|-(\Lambda_{\mathcal{L}}^0\otimes K)\overline{\delta}(t)\|\leq\alpha_1\|\overline{\delta}(t)\|$.
By Assumption \ref{As1} and the $C_r$ inequality, we know that there exists a positive constant $\alpha_2$, such that
$\Big\|\sum_{i=1}^{N}\sum_{j=1}^{N}a_{ij}[\phi^{T}(I_N-J_N)\eta_{N,i}\otimes(Kf_{ji}(\delta_j(t)-\delta_i(t)))]\Big\|\leq \alpha_2\|\overline{\delta}(t)\|$.
Then by \cite[Theorem 4.2]{Mao1997},
we know that  (\ref{controlprotocolexplict}) is an asymptotically unbiased a. s. average-consensus protocol. \qed

\textbf{Proof of Theorem \ref{MSstability}}:
Applying Lemma \ref{linearclosedloopsystemadd1} and the Ito formula gives
$d\|\overline{\delta}(t)\|^2$$\leq$$-\lambda_{\min}(\Psi_{K})$$\|\overline{\delta}(t)\|^2$$+2\sum_{i, j=1}^N$$\overline{\delta}^T(t)[(\phi^TB_{ij}\phi)$ $\otimes$$(K\Sigma_{ji})]\overline{\delta}(t)$$dw_{ji}(t)$,
and $d\|\overline{\delta}(t)\|^2$$\geq$$-\lambda_{\max}(\Psi_{K})$$\|\overline{\delta}(t)\|^2$
$+2\sum_{i,j=1}^N$$\overline{\delta}^T(t)$$[(\phi^TB_{ij}\phi)$$\otimes$$(K\Sigma_{ji})]$$\overline{\delta}(t)dw_{ji}(t)$.
This together with the comparison theorem gives (\ref{meansquarerate}).
If $\Psi_{K}$ is positive definite, then $E[\|\delta(t)\|^2]\to0$.
Also, similar to Theorem \ref{sufficientconlemma}, we have $\lim_{t\to\infty}E[\|x_i(t)-x^*\|^2]=0$, where
$x^*=\frac{1}{N}\mathbf{1}^{T}x(0)+\frac{1}{N}\sum_{i, j=1}^{N}\int_{0}^{\infty}(\mathbf{1}^{T}B_{ij}\otimes K\Sigma_{ji})\delta(t) dw_{ji}(t)$.
By Lemma \ref{bijtechlemmaadd1} and the definitions of $\tilde{\delta}(t)$ and $\overline{\delta}(t)$, applying (\ref{meansquarerate}) gives that \ban
&&\mathbb{E}\Big\|x^{*}-\frac{1}{N}\sum_{j=1}^{N}x_{j}(0)\Big\|^2\cr
&=&\frac{1}{N^2}\sum_{i=1}^N\sum_{j=1}^N\mathbb{E}\int^\infty_0\delta^T(t)(B^T_{ij}\mathbf{1}\mathbf{1}^TB_{ij}\otimes\Sigma^T_{ji}K^TK\Sigma_{ji})\delta(t)dt\cr
&=&\frac{1}{N(N-1)}\sum_{i=1}^N\sum_{j=1}^N\mathbb{E}\int^\infty_0\overline{\delta}^T(t)(\phi^TB^T_{ij}\phi\phi^TB_{ij}\phi\cr
&&\otimes\Sigma^T_{ji}K^TK\Sigma_{ji})\overline{\delta}(t)dt\cr
&\leq&\frac{\lambda_{\max}(\Phi_K)}{N(N-1)}\int^\infty_0\|\overline{\delta}(0)\|^2e^{-\lambda_{\min}(\Psi_K)t}dt\cr
&=&\frac{\lambda_{\max}(\Phi_K)}{N(N-1)\lambda_{\min}(\Psi_K)}\|\delta(0)\|^2,
\ean
which implies (\ref{meanconver2add2}). Then Similar to Theorem \ref{sufficientconlemma}, we know that
(\ref{controlprotocolexplict}) is a m. s. and
a. s. average-consensus protocol. \qed

\textbf{Proof of Theorem \ref{MSstability2}}:
The ``if'' part follows directly from Theorem \ref{sufficientconlemma}.
By the definition of $B_{ij}$ and $\phi$, we have
$\sum_{i, j=1}^N\phi^TB^T_{ij}\phi\phi^TB_{ij}\phi=\frac{2(N-1)}{N}\phi^T\mathcal{L}\phi=\frac{2(N-1)}{N}\Lambda^0_{\mathcal{L}}$.
This together with $K=kI_n$ leads to
$-\Psi_{K}\geq(2k(\frac{k\underline{\sigma}^2(N-1)}{N}-1)\Lambda^0_{\mathcal{L}})\otimes I_{n}$.
Then similarly to Theorem \ref{MSstability}, we have
$d\|\overline{\delta}(t)\|^2\geq2k(\frac{k\underline{\sigma}^2(N-1)}{N}-1)\lambda_2(\mathcal{L})\|\overline{\delta}(t)\|^2+2\sum_{i,j=1}^N\overline{\delta}^T(t)[(\phi^TB_{ij}\phi)\otimes(K\Sigma_{ji})]\overline{\delta}(t)dw_{ji}(t)$,
which imply the ``only if'' part. \qed

%
\textbf{Proof of Theorem \ref{Linalmoststability}}:
By Lemmas \ref{linearclosedloopsystemadd1} and \ref{ORIG}, applying the It\^{o} formula to $\log\|\overline{\delta}(t)\|^2$ gives
$d\log\|\overline{\delta}(t)\|^2
\leq-\mu dt+\frac{2}{\|\overline{\delta}(t)\|^2}\sum_{i,j=1}^N\overline{\delta}^T(t)[(\phi^TB_{ij}\phi)\otimes(K\Sigma_{ji})]\overline{\delta}(t)dw_{ji}(t)$.
Therefore, it follows from the definition of $\mu$ that
\bna
\label{LinEq07}
\frac{2\log\|\overline{\delta}(t)\|}{t}\leq \frac{2\log\|\overline{\delta}(0)\|}{t}-\mu t+\frac{M(t)}{t},
\ena
where
\ban
M(t)=\frac{2}{\|\overline{\delta}(t)\|^2}\sum_{i,j=1}^N\overline{\delta}^T(t)[(\phi^TB_{ij}\phi)\otimes(K\Sigma_{ji})]\overline{\delta}(t)dw_{ji}(t)
\ean
is a local martingale with $M(0)=0$ and the quadratic variations (\cite{Mao1997})
$\frac{\langle M, M\rangle_t}{t}=\sum_{i,j=1}^N\hat{\lambda}_{\max}^2((\phi^TB_{ij}\phi)\otimes(K\Sigma_{ji})+(\phi^TB^T_{ij}\phi)\otimes(\Sigma^T_{ji}K^T))<\infty$.
Applying the law of large number gives
$\lim_{t\rightarrow\infty}\frac{M(t)}{t}=0$, \ a.s.,
which together with (\ref{LinEq07}) gives
$\overline{\lim}_{t\to\infty}\frac{\log\|\delta(t)\|}{t}\leq -\frac{\mu}{2}<0$ a.s.
Then similar to Theorems \ref{MSstability} and
\ref{sufficientconlemma}, we know that the protocol (\ref{controlprotocolexplict}) is an  a. s. average-consensus protocol. \qed


\textbf{Proof of Theorem \ref{symmetricconvergencerate}:}
From Lemma \ref{consensuerrorexplicitsolutionsymmetric}, noting that $A_{\mathcal{L}}(K)$, $M_{\mathcal{L},K}(t)$ and
$\exp\{-A_{\mathcal{L}}(K)t$$+M_{\mathcal{L},K}(t)\}$ are all symmetric matrix and the eigenvalues of
$\exp\{-A_{\mathcal{L}}(K)t$ $+M_{\mathcal{L},K}(t)\}$ are all nonnegative, we know that
$\|\overline{\delta}(t)\|
\leq\exp\{-\lambda_{min}(A_{\mathcal{L}}(K))t+\lambda_{max}(M_{\mathcal{L},K}(t))\}\|\overline{\delta}(0)\|$,
which gives
\bna
\label{logdeltaestiamteaddaddadd1}
&&\frac{\log\|\overline{\delta}(t)\|+\lambda_{min}(A_{\mathcal{L}}(K))t}{\sqrt{2t\log\log t}}\cr
&\leq&\frac{\lambda_{max}(M_{\mathcal{L},K}(t))}{\sqrt{2t\log\log t}}+\frac{\log\|\overline{\delta}(0)\|}{\sqrt{2t\log\log t}}.
\ena
Thus, by the Law of the Iterated Logarithm of Brownian motions, we have (\ref{upperboundrate}).
Similarly,
we have
$\|\overline{\delta}(t)\|
\geq\exp\{-\lambda_{max}(A_{\mathcal{L}}(K))t+\lambda_{min}(M_{\mathcal{L},K}(t))\}\|\overline{\delta}(0)\|$.
From above and the laws of the iterated logarithm of Brownian motions,  similar to (\ref{logdeltaestiamteaddaddadd1}), we have
(\ref{lowerboundrateadd2}).
\qed

\end{document}